\documentclass[10pt, draft, showpacs, amsmath, amssymb, prd, twocolumn]{revtex4}

\def\a{\alpha}
\def\b{\beta}
\def\g{\gamma}
\def\d{\delta}
\def\ep{\varepsilon}

\begin{document}

\title{Caroll-Field-Jackiw electrodynamics in the pre-metric framework}

\author{Yakov Itin}

\affiliation{Institute of Mathematics, Hebrew University of
  Jerusalem and Jerusalem College of Engineering, Jerusalem 91904,
  Israel, email: {\tt itin@math.huji.ac.il}}

\begin{abstract} 
We analyze the Carroll-Field-Jackiw  (CFJ) modification of electrodynamics  reformulated as the ordinary Maxwell theory with an additional special axion field. In this form, the CFJ model appears as a special case of the pre-metric approach recently developed by Hehl and Obukhov. 
 This embedding turns out to be non-trivial. Particularly, the pre-metric energy-momentum tensor does not depend on the axion. This is in contrast to the CFJ energy-momentum tensor which involves the axion addition explicitly. We show that the relation between these two quantities is similar to the correspondence between the Noether conserved tensor and the Hilbert symmetric tensor. As a result the CFJ energy-momentum tensor appears as the unique conserved closure of the pre-metric one. 
Another problem is in the description  of the birefringence effect, which in the pre-metric framework does not depend on the axion.  The  comparison with the CFJ model shows that the corresponding wave propagation (Fresnel) equation has to be extended by a derivative term, which is non zero for the axion field. In this way, the CFJ birefringence effect is derived in the metric-free approach. 
Consequently the Lorentz and CPT violating models can be embedded without contradictions in the pre-metric approach to electrodynamics. This correspondence can be useful for  both constructions.
\end{abstract}


\date{\today}
\maketitle


\section{Introduction --- the CFJ-model}
The Carroll-Field-Jackiw (CFJ) modification of electrodynamics was formulated \cite{Carroll:vb} with a view to examine the possibility of Lorentz and PCT violations in  Maxwell's electrodynamics. 
The model predicts the  rotation of the plane of polarization of radiation from distance galaxies, an effect which is not observed \cite{Carroll:vb}. 
However the original construction gives some general framework to treat Lorentz and PCT violations in field theories. 
Particularly, the similar ideas appear in the Lorentz violating extensions of the standard model \cite{Kostelecky:2003fs}, in the models with the spacetime variation of the coupling constants \cite{Kostelecky:2002ca}, in the Chern-Simon extension of GR  \cite{Jackiw:2003pm}, \cite{Jackiw:2003ri}, and in various other contexts.

In the present paper we show that the original CFJ-model  can be viewed as a specific case of the pre-metric framework of electrodynamics recently developed by Hehl and Obukhov, see \cite{Obukhov:1999ug}, \cite{Hehl:2002hr}, \cite{book}. 

The notations in \cite{Jackiw:2003pm} and \cite{Jackiw:2003ri} are slightly different from those originally used in \cite{Carroll:vb}. Thus we start with a brief account of the CFJ-electrodynamics.

The first field equation for the electromagnetic field (a 2-nd order antisymmetric tensor) $F_{ab}$ is postulated to be the same as in the ordinary electrodynamics: 
\begin{equation}\label{intr1}
\partial_a\,^*F^{ab}=0\,, \qquad ^*F^{ab}=\frac 12\varepsilon^{abcd}F_{cd}\,.
\end{equation}
Consequently, the potential $A_a$ appears in the ordinary form  $F_{ab}=\partial_a A_b-\partial_b A_a$. 
The equation (\ref{intr1})  does not involve the metric of  spacetime. 

The second Maxwell equation 
\begin{equation}\label{intr3}
\partial_a \, F^{ab}=J^{b}\,, \qquad F^{ab}=g^{am}g^{bn}F_{mn}\,
\end{equation}
involves the metric tensor explicitly. In the CFJ-model, it is modified to 
\begin{equation}\label{intr4}
\partial_a \, F^{ab}+v_a \,^*F^{ab}= J^{b}\,,
\end{equation}
 where $v_a$ is a covector. 
The  electromagnetic current $J^{b}$ is conserved, $\partial_b\, J^{b}=0$, provided the covector $v_a$ satisfies 
\begin{equation}\label{intr6}
\partial_a\, v_{b}-\partial_b\, v_{a}=0\,.
\end{equation}

The Lagrangian associated with the modified field equation (\ref{intr4}) involves the Chern-Simon-like term in addition to the standard Maxwell Lagrangian    
\begin{equation}\label{intr7}
L=-\frac 14 F_{ab}F^{ab}+\frac 12 v_a A_b\,^*F^{ab}-J^a A_a\,. 
\end{equation}
This expression is gauge invariant provided (\ref{intr6}) holds. 

The energy-momentum tensor for the electromagnetic field, in the CFJ theory, is given in the form \cite{Carroll:vb}
\begin{equation}\label{intr8}
{\Theta^a}_b=-F^{ac}F_{bc}+\frac 14 \d^a_bF^{cd}F_{cd}+\frac 12 {v_b}\,^*F^{ac}A_c\,.
\end{equation}
The divergence of this tensor is equal to the standard Lorentz force expression
\begin{equation}\label{intr9}
\partial_a{\Theta_b}^a=J^a F_{ab}\,,
\end{equation}
provided (\ref{intr6}) is restricted to 
\begin{equation}\label{intr9x}
\partial_a v_b=0\,.
\end{equation}
Consequently, in the absence of sources, (\ref{intr8}) is conserved by virtue of the field equations. 

The Chern-Simon-type addition breaks the symmetry of the energy-momentum tensor (\ref{intr8}) and its positive definiteness.  
This fact is treated in \cite{Carroll:vb} as an indication of the absence of the Lorentz invariance.  Observe also that ${\Theta_b}^a$ has a non zero trace ${\Theta_{a}}^{a}=(1/2) {v_a}\,^*F^{ac}A_c$.

The appearance  of a ``fixed'' (in the sense of (\ref{intr9x})) covector $v_a$  in the field equation  (\ref{intr3}) and in the Lagrangian (\ref{intr7}) is treated in \cite{Carroll:vb} as a breaking of the Lorentz invariance of the theory. Moreover, if $v_a$ is considered as a covector (not a pseudo-covector), the parity symmetry is lost too. To preserve  $SO(3)$ space invariance, the covector $v_a$ is taken in the form 
\begin{equation}\label{intr9xx}
v_a=(\mu,0,0,0)\,,
\end{equation}
where  $\mu$ is a constant.

A physical consequence of the modified electrodynamics appears as a certain variation of the dispersion law. For a plane monochromic wave it takes the form \cite{Carroll:vb}
\begin{equation}\label{intr10}
w=\sqrt{|{\bf k}|^2\pm \mu|{\bf k}|}\approx |{\bf k}|\pm \frac {\mu}2\,.
\end{equation}
This birefringence of the vacuum generates a Faraday like rotation on polarized light. Actual astronomical measurements impose an upper bound on the parameter $\mu$. 

 As it was already mentioned in \cite{Carroll:vb}, the CFJ model can be equivalently reformulated as an ordinary electrodynamics with  additional axion field. This new formulation is not complete because the corresponding axion field is non-dynamical. However, two equivalent representations of the model have to give the same physical consequences. The non-dynamical axion field naturally appears in the pre-metric approach to electrodynamics recently developed by Hehl an Obukhov \cite{Obukhov:1999ug}, \cite{Hehl:2002hr}, \cite{book}.  
The comparison of this framework to the CFJ-model raises two  problems: 
 
 (i)  The energy-momentum tensor of pre-metric electrodynamics does not depend on the axion field.  The CFJ energy-momentum tensor involves the axion field.  How are these two energy-momentum tensors  related to one  other? 
 
(ii) The wave propagation, in the pre-metric electrodynamics, is managed by an equation of the 4th order (Fresnel equation). The axion field drops out from this expression \cite{book}. How does the CFJ birefringence of the vacuum (\ref{intr10}) appear in this context? 

 The pre-metric approach is manifestly covariant so it  would be interesting to find out how the Lorentz and PCT violating model is embedded in this covariant model. 
Observe that, for a covariant covector $v_a$,  the field equations, the Lagrangian and the energy-momentum tensor of the CFJ model are Lorentz invariant. The covector $v_a$ can always be taken in the form (\ref{intr9xx}). This choice declares a sub-class of coordinate systems that are related by $SO(3)$ transformations. The Lorentz invariance is actually broken, in the CFJ model, only when the specific choice (\ref{intr9xx}) of the covector $v_a$  is assumed to hold in an arbitrary coordinate system. Thus an additional problem occurs: 
 
 (iii) How do these two alternative descriptions of the covector $v_a$ influence the dispersion law? In other words: Is the  birefringence effect  really an indication of the Lorentz and CPT symmetry breaking? 

 


Our goal is to discuss the indicated problems. The plan of the paper is as follows: 
We start with a brief account of pre-metric electrodynamics. Although the exterior form technique is more convenient in this construction, we use the tensorial form which allows a straightforward comparison with the CFJ-model.  

Our main results are given in the next two sections. By introducing the CFJ excitation and the corresponding constitutive tensor, we reformulate the CFJ-model in the pre-metric form.  The pre-metric energy-momentum tensor turns to be of the Hilbert type, while the CFJ expression is a Noether conserved quantity.  Moreover, we derive the CFJ quantity by means of the conserved closure of the pre-metric tensor. In an opposite way, the pre-metric tensor is the symmetric part of the CFJ one. 

In the  fourth section, we deal with  the effects of the light propagation. By comparison with the CFJ model, we show that the pre-metric Fresnel equation \cite{Hehl:2002hr}, \cite{book}  has to be supplemented with the terms  involving the derivatives of the constitutive tensor. For the models of the CFJ type with the fixed metric and the variable axion part, we derive the explicit form of the additional term.  The corresponding Fresnel equation appears to be of the 4-th order as in the case of a constant constitutive tensor.  


\section{The pre-metric approach to electrodynamics}
In the pre-metric  electrodynamics \cite{Post}, \cite{book} one starts with the spacetime considered as a $4$ dimensional differential manifold without additional structures as metric or connection. The conservation of the electro-magnetic current 
is treated as a basic fundamental fact. When  necessary constraints on the topology of the spacetime are applied, this conservation law results in the existence of the electromagnetic excitation $H^{ab}$ --- an antisymmetric 2-nd order tensor 
\begin{equation}\label{mfr1}
\partial_b\, J^{b}=0\quad {\rm {yields}} \quad\partial_a H^{ab}= J^b\,.
\end{equation}
The quantities $J^a$ and $H^{ab}$ are considered as twisted (odd) tensors (differential forms). It means that they change their signs under a map of  spacetime with a negative determinant, for instance, under $P$ or $T$ transformations.

The second basic fact assumed in the pre-metric framework is  the Lorentz law of interaction between the charges and the electromagnetic field  
\begin{equation}\label{mfr2}
f_a=J^b F_{ab}\,.
\end{equation}
This law is viewed as an operational definition of the  electromagnetic strength $F_{ab}$. This is an untwisted (even) tensor, which does not change under the maps  of  spacetime with a negative determinant.  
The first Maxwell field equation 
\begin{equation*}
\partial_a\,^*F^{ab}=0
\end{equation*}
is treated as an expression of the magnetic flux conservation law. In other wards, the absence of magnetic monopoles is considered as a fundamental fact. 

The  excitation and the electromagnetic strength can be related, in general, by an operator
\begin{equation}\label{mfr1x}
H^{ab}=\kappa^{ab}(F_{cd})\,.
\end{equation}
The restriction of this equality to a linear, homogeneous, local relation yields
\begin{equation}\label{mfr2x}
H^{ab}=\chi^{abcd}F_{cd}\,,
\end{equation}
where $\chi^{abcd}$ is the constitutive tensor, with the symmetries
\begin{equation}\label{mfr3x}
\chi^{abcd}=-\chi^{bacd}=-\chi^{abdc}\,.
\end{equation}
The irreducibly decomposition of this tensor, under the group of linear transformations, involves three independent pieces 
\begin{equation}\label{mfr4}
\chi^{abcd}={}^{\tt (1)}\chi^{abcd}+{}^{\tt (2)}\chi^{abcd}+{}^{\tt (3)}\chi^{abcd}\,.
\end{equation}
The axion and the skewon parts are defined as  \cite{rem0}
\begin{equation}\label{mfr5}
{}^{\tt (3)}\chi^{abcd}=\chi^{[abcd]}\,,\quad {}^{\tt (2)}\chi^{abcd}=\frac 12(\chi^{abcd}-\chi^{cdab})
\end{equation} 
while the principal part is 
\begin{equation}\label{mfr6}
{}^{\tt (1)}\chi^{abcd}=\chi^{abcd}-{}^{\tt (2)}\chi^{abcd}-{}^{\tt (3)}\chi^{abcd}\,.
\end{equation} 

The Lagrangian, in  pre-metric electrodynamics, may be taken in the form 
\begin{equation}\label{mfr9}
{\cal L}=-\frac 14\, F_{ab}H^{ab}-J^a A_a\,.
\end{equation} 
The skewon part $^{\tt (2)}\chi^{abcd}$ is not involved in (\ref{mfr9}). The variation relative to the potential field $A_a$ yields the field equation (\ref{mfr1}), see \cite{Itin:2003jp}. 

The energy-momentum tensor, in the pre-metric electrodynamics, is postulated as 
\begin{equation}\label{mfr10}
{T_b}^{a}=-F_{bm}H^{am}+\frac 14 \delta_b^a F_{mn}H^{mn}\,. 
\end{equation} 
This tensor is traceless for an arbitrary constitutive tensor. Its divergence equals to the Lorentz force plus an additional term that depends on the derivatives of $\chi^{abcd}$. 
The axion part ${}^{(3)}\chi^{abcd}$ is also not involved  in (\ref{mfr10}) \cite{book}. 

The wave propagation in this pre-metric framework is managed by a 4th order equation for the wave-covector $q_a$
\begin{equation}\label{mfr11}
{\mathcal G}^{abcd}q_a q_b q_c q_d=0\,,
\end{equation} 
where 
\begin{equation}\label{mfr12}
{\mathcal G}^{abcd}=\frac 1 {4!}\epsilon_{mnpq}\epsilon_{rstu}\chi^{mnr(a}\chi^{b|ps|c}\chi^{d)qtu}
\end{equation} 
is the  tensor density of the weight $+1$. The axion part drops out  also from this tensor \cite{book}. 
\section{CFJ embedded in metric free framework}
\subsection{The CFJ constitutive tensor}
We start with embedding  standard electrodynamics in the pre-metric framework. Comparing the equations (\ref{intr3}) and (\ref{mfr1}) we see that the Maxwell excitation $H^{ab}$ has to be equal to $F^{ab}$. Take also into account that the current $J^a$ and the excitation $H^{ab}$ have to be treated as twisted (odd) tensors, while $F^{ab}$ is untwisted. Consequently, the Maxwell excitation is 
\begin{equation}\label{emb00}
^{\tt (Max)}H^{ab}=\sqrt{-g}g^{ac}g^{bd}F_{cd}\,.
\end{equation} 
Hence  the Maxwell constitutive tensor 
\begin{equation}\label{emb0}
^{\tt (Max)} \chi^{abcd}=\frac 12 \sqrt{-g}\left(g^{ac}g^{bd}-g^{bc}g^{ad}\right)\,,
\end{equation}
involves only the principal part. 

The CFJ modified field equation (\ref{intr4}) involves the term proportional to $F^{ab}$ itself, i.e., to the first order derivatives of the potential $A_a$. Thus the CFJ excitation can be constructed as a type of  integral operator.  Although the general case of a linear operator relation between $H_{ab}$ and $F_{ab}$ was mentioned above (\ref{mfr1x}),  it is  preferable to deal with an ordinary local tensor expression. For this, we   rewrite  (\ref{intr4}) in an equivalent form. The condition (\ref{intr6}) is satisfied by a covector $v_a$ equal to the gradient of an arbitrary function 
\begin{equation}\label{emb1}
v_a=\partial_a \theta\,, \qquad \theta=\theta(x^a)\,.
\end{equation}
With this redefinition, the field equation (\ref{intr4}) is equivalent to
\begin{equation}\label{emb3}
\partial_a \left(F^{ab}+\theta\,^*F^{ab}\right)=J^{b}\,,
\end{equation}
provided the field equation (\ref{intr1}). 
Comparing  with (\ref{mfr1}) we derive the CFJ electromagnetic excitation
\begin{equation}\label{emb4}
H^{ab}=\sqrt{-g}F^{ab}+\theta\,^*F^{ab}\,,
\end{equation}
i.e., the constitutive tensor in  CFJ electrodynamics involves an axion in addition to the principal part 
\begin{equation}\label{emb5}
^{\tt (CFJ)} \chi^{abcd}=\frac 12\, \sqrt{-g}\left(g^{ac}g^{bd}-g^{bc}g^{ad}\right)+\frac 12\,\theta\,\varepsilon^{abcd}\,.
\end{equation}
Observe that the function $\theta(x)$ is arbitrary in (\ref{emb4},\ref{emb5}), thus both expressions are Lorentz invariant. As for the parity invariance, for $\theta(x)$  a scalar, parity symmetry is loosed, while, for $\theta(x)$ is a pseudo-scalar (twisted scalar), it is preserved. We will consider these two cases simultaneously. It means, that we allow the constitutive tensor  to have a non-homogeneous parity although, in  pre-metric electrodynamics, it is assumed  to be twisted.  

\subsection{The CFJ Lagrangian}
Substituting the CFJ excitation (\ref{emb4}) into the pre-metric Lagrangian  (\ref{mfr9}) we obtain
\begin{equation}\label{emb6}
{\cal L}=-\frac 14\, F_{ab}F^{ab}\sqrt{-g}-\frac 14\theta F_{ab}\,^*F^{ab}-J^a A_a\,.
\end{equation} 
Observe that
\begin{eqnarray}\label{emb7}
\frac 14\,\theta F_{ab}\,^*F^{ab}&=&-\frac 12\,v_a A_b\,^*F^{ab}-\frac 12\,\theta A_b\,\partial_a\,^*F^{ab}+\nonumber\\
&&\frac 12\,\partial_a\left(\theta A_b\,^*F^{ab}\right)\,.
\end{eqnarray}
The second term vanishes,  provided the field equation (\ref{intr1}). 
Thus the pre-metric Lagrangian (\ref{mfr9}) is equivalent to the CFJ one (\ref{intr7}), up to a total derivative.
\subsection{The CFJ energy-momentum tensor}
Substituting (\ref{emb4}) in the energy-momentum tensor of  metric-free electrodynamics (\ref{mfr10}) we obtain the ordinary Maxwell energy-momentum tensor  
\begin{equation}\label{emb13}
{}^{({\tt Max})}{T_b}^a=-F^{ac}F_{bc}+\frac 14 \d^a_bF^{cd}F_{cd}
\end{equation}
since the axion term does not give  any addition \cite{rem0}. 
This expression is, in fact,  the Hilbert energy-momentum tensor. Indeed,  since the Chern-Simon-like term is independent on the metric, $T^{ab}={\delta L}/{\d g_{ab}}={\delta }{\mathcal L}/{\d g_{ab}}\,. $

Certainly, (\ref{emb13}) is  traceless, symmetric and positive defined. 
Its divergence, however, is not equal to the Lorentz force. In order to compare this expression to the CFJ tensor (\ref{intr8}), we consider the divergence of the general expression (\ref{mfr10}). It takes the form \cite{book} 
\begin{equation}\label{emb15}
\partial_a {T_b}^a= f_b+X_b\,,
\end{equation}
where 
\begin{equation}\label{emb16}
X_b=\frac 14\left(H^{ij}\partial_b F_{ij}-F_{ij}\partial_b H^{ij}\right)\,.
\end{equation}
Substituting here the constitutive relation (\ref{mfr2x}), we obtain the reduction to two terms of different origin $X_b=Y_b+Z_b$, with 
\begin{equation}\label{emb17}
Y_b=-\frac 14F_{ij}F_{\rho\sigma}\partial_b\left({}^{(1)}\chi^{ijkl}+{}^{(3)}\chi^{ijkl}\right)\,,
\end{equation}
and
\begin{equation}\label{emb18}
Z_b=\frac 14\,\,{}^{(2)}\chi^{ijkl}F_{ij}\partial_b F_{kl}\,.
\end{equation}
In the second term, $Z_b$, the derivatives of the electromagnetic field are involved, so it is of the same fashion as the Lorentz force itself. Thus $Z_b$  gives a type of a ``hard violation'' of the conservation law. This fact is rather natural, since the skewon term is not involved in the Lagrangian. 
In the first term, $Y_b$, only the derivatives of the constitutive tensor are involved. This term is not zero  for a spacetime with a variable metric even without the axion term modification. This is a type of a ``soft violation'' of the conservation law. It  is usually treated by the change of the partial derivative to the covariant one.  

In the CFJ model we deal with a constant principal part and a variable axion part (\ref{emb5}). Moreover, the axion part does not involve the energy-momentum tensor (\ref{mfr10}). Consequently the relation (\ref{emb15}) takes the form 
\begin{equation}\label{emb19}
\partial_a\,{}^{({\tt Max})}{T_b}^a= f_b-\frac 14F_{mn}\,^*F^{mn}\partial_b\theta\,.
\end{equation}
This relation is  equivalent to 
\begin{equation}\label{emb20}
\partial_a\left({}^{({\tt Max})}{T_b}^a+\frac 12 \,^*F^{ac}A_c\partial_b\theta\right)=f_b\,
\end{equation}
 provided the field equation (\ref{intr1}) and the condition (\ref{intr9x}). 
The expression in  brackets coincides with the CFJ energy-momentum tensor (\ref{intr8}). Thus this tensor can be viewed as a conserved closure of the pre-metric energy-momentum tensor. This extension is unique up to a total derivative. 

A more meaningful description of (\ref{intr8}) can be given in the framework of the Noether procedure. Indeed, this expression is derivable from the Lagrangian (\ref{intr7}) by the standard formula for the canonical energy-momentum tensor. 
Thus  (\ref{intr8})  can be interpreted as an energy-momentum tensor of a system of two interacted fields --- the electromagnetic field $F_{ab}$ and the non-dynamical field $v_a$, see \cite{Itin:2003jp}.  The additional term can be viewed as the energy of interaction of the electromagnetic field with an ``infinite sea'' of the covector field $v_a$. It is rather natural that the total expression is not positive defined and non-symmetric. These features of the energy-momentum tensor cannot be seen, however, as an indication of the Lorentz or parity symmetry breaking. They appear even in the case of the covector field restricted by a Lorentz invariant condition (\ref{intr9x}). 
 
 The quantity (\ref{intr8}) shares the well known  properties of  canonical energy-momentum tensors. Particularly, it is gauge invariant only up to a total derivative. It is a general property of the Noether tensors derived from the Lagrangians which are gauge invariant only up to a total derivative \cite{Itin:2003jp}. Although the total derivative is not important in most situations, it prevents (\ref{intr8}) from being used as a source in Einstein's gravity equation. 
\section{Light propagation}
In the framework of the pre-metric approach, the light propagation is managed by the 4th order equation (\ref{mfr11}). The axion part of the constitutive tensor does not alter this equation \cite{book}.  Consequently, for the CFJ constitutive tensor (\ref{emb5}), the equation (\ref{mfr11}) does not give any birefringence effect at all. This is in a contradiction with the CFJ dispersion law (\ref{intr10}).  In fact,  the equation (\ref{mfr11}) is derived in the geometrical optics approximation. Thus the derivatives of the constitutive tensor are neglected.  However, the CFJ birefringence effect is proportional to $\mu^2$, i.e., to the square of the first order derivative of $\chi^{abcd}$.  The CFJ dispersion law uses the exact plane wave solution, while (\ref{mfr11}) is based on the geometrical optics limit.  Thus, in order to have a correspondence between two formulas, the expression (\ref{mfr12}) has to be supplied with a certain  correction term. 
 
In fact we have here two different types of the birefringence effects: (i) The pre-metric birefringence is generated by an algebraic structure of the constitutive tensor. (ii) The  CFJ birefringence effect is generated by derivatives of the constitutive tensor.

For description of the light propagation, we consider  the wave-type ansatz 
\begin{equation}\label{lp1}
F_{ab}=f_{ab}\,e^{i\varphi}\,, 
\end{equation}
 where $\varphi=\varphi(x^a)$ while $f_{ab}$ is a constant tensor. 
We denote the wave covector as $q_a=\partial_a\varphi$.  
For the ansatz (\ref{lp1}), the tensor of excitation (\ref{mfr1x}) is  
\begin{equation}\label{lp2}
H^{ab}=\chi^{abcd}f_{cd}\,e^{i\varphi}\,. 
\end{equation}
 Observe, that, in general, the amplitude of $H_{ab}$ is not a constant tensor, even if $f_{ab}$, the amplitude of $F_{ab}$, is a constant.  
 
Substituting  (\ref{lp1}, \ref{lp2}) into the field equations (\ref{intr1}) and  (\ref{mfr1}) and putting to zero the current vector, we obtain a system of 8 linear equations 
\begin{eqnarray}\label{lp3}
&&\ep^{abcd}q_bf_{cd}=0\\
\label{lp4}
&&\Big(\chi^{abcd}q_b-i{\chi^{abcd}}_{,a}\Big)f_{cd}=0
\end{eqnarray}
for 6 independent variables $f_{ab}$. For a constant constitutive tensor, this system coincides with the corresponding system of \cite{Hehl:2002hr}, which was derived by means of the Hadamard method. 
The equations (\ref{lp3}) and (\ref{lp4}) are linearly dependent. Indeed, the contraction of (\ref{lp3}) with the covector $q_a$ is identically zero. 
The solution of (\ref{lp3}) may be written as 
\begin{equation}\label{lp5}
f_{ab}=q_aa_b-q_ba_a\,, 
\end{equation}
where $a_a$ is an arbitrary covector. It is defined only up to an arbitrary shift $a_a\to a_a+\lambda q_a$, which  corresponds to the gauge transformations of the potential. 
In order to fix this ``gauge'' freedom, we use the covector in the form of  Tamm's ansatz \cite{Hehl:2002hr}
\begin{equation}\label{lp6}
a_a=\frac{a_0}{q_0}q_a+l_a\,.
\end{equation}
The  new covector $l_a$  has only 3 independent components $l_a=(0,l_\a), \a=1,2,3$. Substituting  (\ref{lp6}) into  (\ref{lp4}) we obtain 4 linear equations 
\begin{equation}\label{lp7}
\Big(\chi^{abc\d}q_bq_c+i{\chi^{abc\d}}_{,b}q_c\Big)l_\d=0\,
\end{equation}
 for 3 independent components $l_\a$. 
Following \cite{Hehl:2002hr}, \cite{book} we introduce a specific frame such that  the wave vector is directed in the positive time direction, i.e.,  $q_a=(1,0,0,0)$.
In $1+3$ decomposition, (\ref{lp7}) reads 
\begin{eqnarray}\label{lp8}
&&{\chi^{0\b0\d}}_{,\b}\,l_\d=0\,,\\
\label{lp9}
&&\left(\chi^{\a 00\d}+i{\chi^{\a\b 0\d}}_{,\b}+i{\chi^{\a0 0\d}}_{,0}\right)l_\d=0\,.
\end{eqnarray}
We still deal with an over-determined system of 4 linear equations. 
  
Consider now the special  case appearing in the CFJ-model. The corresponding constitutive tensor involves the constant principal part ${}^{(1)}\chi^{abcd}$ and the variable axion part ${}^{(3)}\chi^{abcd}=\theta(x^\mu)\ep^{abcd}$.
Consequently, the equation (\ref{lp8}) is satisfied identically and we remain with  the system (\ref{lp9}) of 3 independent equations ($v_\b=\partial_\b \theta$)
\begin{equation}\label{lp10}
M^{\a\d}a_\d=0\,, \quad {\rm with}\quad M^{\a\d}={}^{(1)}\chi^{\a 00\d}+iv_\b\ep^{\a\b 0\d}\,.
\end{equation}
A nontrivial solution to this homogeneous linear system appears only for ${\rm det}\, M^{\a\d}=0$, i.e., 
\begin{equation}\label{lp11}
\frac 1{3!}\ep_{\a\b\g}\ep_{\mu\nu\rho}M^{\a\mu}M^{\b\nu}M^{\g\rho}=0\,.
\end{equation}
We substitute here (\ref{lp10}) and observe that the imaginary part vanishes because of symmetries (the skewon part is absent in the CFJ model). 
Consequently we come to the equation 
\begin{equation}\label{lp12}
\frac 1{3!} \ep_{\a\b\g}\ep_{\mu\nu\rho}\chi^{\a00\mu}\chi^{\b00\nu}\chi^{\g00\rho}-
v_\a v_\b\chi^{\a00\b}=0\,.
\end{equation}
Following the procedure given in \cite{Hehl:2002hr}, \cite{book} we rewrite this equation in the covariant form 
\begin{equation}\label{lp13}
{\mathcal{G}}^{abcd}q_aq_bq_cq_d-\chi^{abcd}v_a v_dq_bq_c=0\,.
\end{equation}
 Substituting here the CFJ constitutive tensor (\ref{emb5}) we obtain 
 \begin{equation}\label{lp14}
 (q_aq^a)^2-(v_aq^a)^2+(v_av^a)(q_bq^b)=0\,,
 \end{equation}
 which coincides with \cite{Jackiw:1998js}. 
 Observe that for a derivation of this relation we only need the covariant condition (\ref{intr6}) which allows to embed the CFJ model in the pre-metric setting. 
 
Finally,  in the rest frame $v_a=(m,0,0,0)$, we substitute  $q_a=(w,{\bf k})$ to derive 
  \begin{equation}\label{lp15}
 (w^2-{\bf k}^2)^2-m^2{\bf k}^2=0\,,
  \end{equation}
which coincides with the CFJ dispersion law (\ref{intr10}). 
 \section{Discussion}

In the case the violations of the Lorentz and the CPT symmetries exist as  electromagnetic  phenomena, it is very possible that they would appear in the form of the CFJ modification.
The reason is that this model preserves the basic features of  ordinary electrodynamics. In particular, (i) the electromagnetic charge is conserved, (ii) the model is gauge invariant, (iii) the divergence of the  energy-momentum is equal to the standard Lorentz force, (iv)  the electromagnetic flux is conserved (absence of monopoles), and (v) the model is derivable from a Lagrangian. 

The birefringence effect is a measurable result of this model.  So, in  the case the birefringence is not observable, one can deduce  that there is no  violation of Lorentz and CPT, as  in \cite{Carroll:vb}. 
The  inverse deduction, in general, is not true. Even if the birefringence effect  was  observable, it could be originated  also in  a corresponding Lorentz and CPT invariant model. 

 The full set of Lorentz violation terms in electrodynamics is considered in \cite{Colladay:1998fq}, \cite{Kostelecky:2002hh}. The corresponding addition in the Lagrangian  is 
 \begin{equation}\label{Kos}
 \Delta L=-\frac 14 (k_F)_{abcd}F^{ab}F^{cd}\,.
 \end{equation}
 For $(k_F)_{abcd}$ treated as a set of coupling constants (not a tensor), this is a CPT even Lorentz violating term.
 Only 10 of the 19 possible terms of the form (\ref{Kos}) generate the birefringence effect. 
The coefficient matrix $(k_F)_{abcd}$ is very similar to the constitutive tensor $\chi^{abcd}$. Thus the modification (\ref{Kos}) is in a straightforward way embedded in the pre-metric framework. 


\medskip
 
\noindent{\it Acknowledgment:\/}  I am grateful to Roman Jackiw, Alan Kostelecky, Friedrich Hehl, and Yuri Obukhov for useful comments. The work has been supported by the Minerva foundation.


\begin{thebibliography}{99}
\bibitem{Carroll:vb}
S.~M.~Carroll, G.~B.~Field and R.~Jackiw,
Phys.\ Rev.\ D {\bf 41}, 1231 (1990).
 
\bibitem{Jackiw:1998js}
R.~Jackiw,
Comments Mod.\ Phys.\ A {\bf 1}, 1 (1999) 
 
\bibitem{Kostelecky:2003fs}
A.~Kostelecky,
arXiv:hep-th/0312310.

\bibitem{Kostelecky:2002ca}
V.~A.~Kostelecky, R.~Lehnert and M.~J.~Perry,
Phys.\ Rev.\ D {\bf 68}, 123511 (2003)

\bibitem{Jackiw:2003pm}
R.~Jackiw and S.~Y.~Pi,
Phys.\ Rev.\ D {\bf 68}, 104012 (2003)

\bibitem{Jackiw:2003ri}
R.~Jackiw,
arXiv:gr-qc/0310115.



\bibitem{Obukhov:1999ug}
Y.~N.~Obukhov and F.~W.~Hehl,
Phys.\ Lett.\ B {\bf 458}, 466 (1999)
[arXiv:gr-qc/9904067].

\bibitem{Obukhov:2002xa}
Y.~N.~Obukhov and G.~F.~Rubilar,
Phys.\ Rev.\ D {\bf 66}, 024042 (2002)
[arXiv:gr-qc/0204028].

\bibitem{Hehl:2002hr}
F.~W.~Hehl, Y.~N.~Obukhov and G.~F.~Rubilar,
Int.\ J.\ Mod.\ Phys.\ A {\bf 17}, 2695 (2002)

\bibitem{book} F.W.~Hehl and Yu.N.~Obukhov, {\it Foundations of
    Classical Electrodynamics: Charge, Flux, and Metric}
  (Birkh\"auser: Boston, MA, 2003).

\bibitem{Post} E.J.\ Post, {\it Formal Structure of Electromagnetics
    -- General Covariance and Electromagnetics} (North Holland:
  Amsterdam, 1962, and Dover: Mineola, New York, 1997).

\bibitem{rem0} The brackets $(a,b,\cdots)$ and $[a,b,\cdots]$ mean correspondingly the symmetrization and the antisymmetrization relative to all included indices. The exclusion of some indices is denoted by vertical lines as in (\ref{mfr12}). In some proofs, we  use the relation $\d_a^{\{b}\varepsilon^{cdef\}}=0$, where the brackets $\{\cdots\}$ mean the cyclic permutation.


\bibitem{Itin:2003jp}
Y.~Itin,
J.\ Phys.\ A {\bf 36}, 8867 (2003)


\bibitem{Colladay:1998fq}
D.~Colladay and V.~A.~Kostelecky,
Phys.\ Rev.\ D {\bf 58}, 116002 (1998)
[arXiv:hep-ph/9809521].

\bibitem{Kostelecky:2002hh}
V.~A.~Kostelecky and M.~Mewes,
Phys.\ Rev.\ D {\bf 66}, 056005 (2002)
[arXiv:hep-ph/0205211].
 
\end{thebibliography}
\end{document}